\documentclass[prl,letterpaper,nobalancelastpage,twocolumn,superscriptaddress,longbibliography]{revtex4-1}

\usepackage{graphicx}
\usepackage{amsmath}
\usepackage{bbold}
\usepackage{amssymb}
\usepackage[english]{babel}
\usepackage{color}
\usepackage[version=4]{mhchem}
\usepackage[hidelinks]{hyperref}
\usepackage{dsfont}

\usepackage{soul}
\newcommand{\Caltech}{Division of Physics, Mathematics and Astronomy, California Institute of Technology, Pasadena, CA 91125, USA}

\newcommand{\Harvard}{Department of Physics, Harvard University, Cambridge, MA 02138, USA}

\renewcommand{\cite}[1]{\mbox{\citep{#1}}}

\addto\captionsenglish{}
\addto\captionsenglish{}

\begin{document}

\title{Dark-state enhanced loading of an optical tweezer array}
\author{Adam L. Shaw}
\author{Pascal Scholl}
\author{Ran Finklestein}
\author{\\Ivaylo S. Madjarov}
\affiliation{\Caltech}
\author{Brandon Grinkemeyer}
\affiliation{\Harvard}
\author{Manuel Endres}\email{mendres@caltech.edu}
\affiliation{\Caltech}

\begin{abstract}
Neutral atoms and molecules trapped in optical tweezers have become a prevalent resource for quantum simulation, computation, and metrology. However, the maximum achievable system sizes of such arrays are often limited by the stochastic nature of loading into optical tweezers, with a typical loading probability of only 50\%. Here we present a species-agnostic method for dark-state enhanced loading (DSEL) based on real-time feedback, long-lived shelving states, and iterated array reloading. We demonstrate this technique with a 95-tweezer array of $^{88}$Sr atoms, achieving a maximum loading probability of 84.02(4)\% and a maximum array size of 91 atoms in one dimension. Our protocol is complementary to, and compatible with, existing schemes for enhanced loading based on direct control over light-assisted collisions, and we predict it can enable close-to-unity filling for arrays of atoms or molecules.
\end{abstract}

\maketitle

\textit{Introduction.}---Arrays of neutral atoms or molecules trapped in optical tweezers are a powerful tool for quantum science, due in large part to the availability of single-particle control~\cite{Kaufman2021,Browaeys2020}. In typical experimental sequences, atoms are loaded into traps, imaged, and then dynamically rearranged~\cite{Endres2016,Barredo2016} into a final configuration with real-time tunability of inter-particle separations. This flexibility allows access to a variety of realizable Hamiltonians such as those featuring dipolar~\cite{Bao2022,Holland2022,Leseleuc2019} or Rydberg~\cite{Bernien2017,Madjarov2020,Browaeys2020} interactions. Tweezer-based quantum simulators show competitive entanglement-generation fidelities~\cite{Choi2023,Madjarov2020,Levine2019}, making scaling such systems a major near term goal for applications in quantum simulation~\cite{Scholl2021,Semeghini2021}, computation~\cite{Madjarov2020,Levine2019,Bluvstein2022,Graham2022}, and metrology~\cite{Madjarov2019,Young2020}.

\begin{figure}[t!]
	\centering
	\includegraphics[width=86mm]{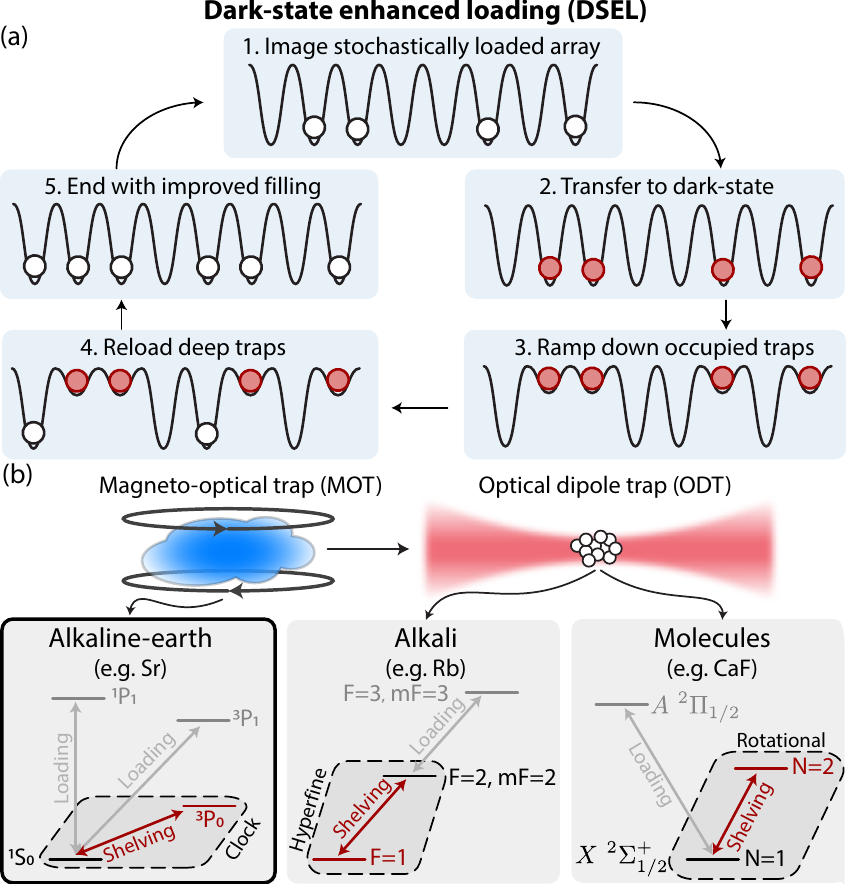}
	\caption{
	(a) A scheme for dark-state enhanced loading (DSEL) of an optical tweezer array. (1) Atoms (or molecules) are initially stochastically loaded into a tweezer array, and imaged to determine their locations. (2) Atoms are transferred to a long lived shelving state which is dark to the loading dynamics, and (3) occupied traps are ramped down in depth to limit reloading. (4) The loading procedure is repeated, stochastically reloading the unoccupied deep traps while leaving the shelved dark-state atoms unaffected, and (5) all atoms are returned to the ground state for an effective increase in filling fraction. The procedure can be repeated for successively higher filling fraction. (b) DSEL is feasible to implement in arrays of alkaline-earth atoms (clock state shelving, e.g. Sr), alkali atoms (hyperfine state shelving, e.g. Rb), or molecules (rotational state shelving, e.g. CaF); the latter two both require tweezers be loaded from an ODT, rather than directly from a MOT. In this work we show an experimental demonstration with the alkaline-earth species $^{88}$Sr.
	}
	\vspace{-0.5cm}
	\label{fig1}
\end{figure}

However, even while array sizes steadily increase through use of higher-power trapping lasers and other technical improvements, efforts to efficiently scale to larger arrays are hindered by imperfect atom loading. In typical experiments, a magneto-optical trap (MOT) or large-waist optical dipole trap (ODT) containing a cloud of atoms is overlapped with a set of tweezer potentials which each capture an ensemble of atoms. Traps are then illuminated to induce light-assisted collisions and pairwise loss~\cite{Schlosser2001,Fung2015,Kaufman2021}, resulting in a typical loading fidelity of 50\% ($\sim$35\% for molecular tweezer arrays~\cite{Anderegg2019,Holland2022}). Through direct control of the collisional process~\cite{Grunzweig2010,Lester2015}, the array-averaged filling has been raised as high as 74-80\% for arrays of alkali species~\cite{Brown2019,Aliyu2021,Angonga2022} and 93\% for Yb~\cite{Jenkins2022}. While tweezer rearrangement can be used to eliminate remaining holes, the success probability and rearrangement time both scale poorly with the number of initial defects in the array~\cite{Endres2016,Barredo2016,Schymik2020}, making methods for improving the initial loading not just impactful in terms of increasing atom number, but also in terms of increasing overall operational fidelity.

\begin{figure}[t!]
	\centering
	\includegraphics[width=86mm]{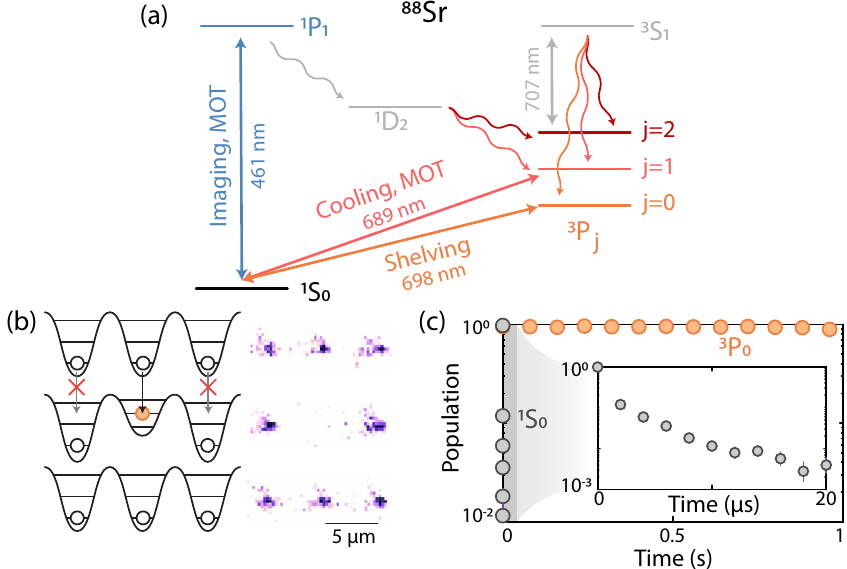}
	\caption{(a) Low-lying state manifold for $^{88}$Sr. Of particular importance is the $^3P_0$ state which is long-lived, and dark to imaging, cooling, and MOT dynamics. (b) Consecutive, single-shot fluorescence images of: (top) all atoms initially in the ground state; (middle) select atoms locally shelved via frequency-selective motional sideband driving to $^3P_0$, and thus hidden from imaging; (bottom) all atoms brought back to the ground state, indicating the survival of the $^3P_0$ atoms. In all images 461, 689, and 707 nm light is illuminating the array. (c) $^3P_0$ state population (orange) at $12\ \mu$K trap depth while the blue and red MOT light is on, showing a 23(4) s lifetime. In contrast, atoms in $^1S_0$ (grey) at this trap depth are quickly heated and expelled from the trap in less than 2 $\mu$s with the MOT light on (inset).
	} 
	\vspace{-0.5cm}
	\label{fig2}
\end{figure}

Here we present a species-agnostic technique, termed Dark-State Enhanced Loading (DSEL), for improving array loading beyond the limits set by light-assisted collisions. We demonstrate this protocol with a 95-tweezer array filled with the alkaline-earth $^{88}$Sr atom, for which a demonstration of enhanced loading has not been shown previously. We achieve a final filling of 84.02(4)\% over the array for an initial loading probability of 45.88(5)\%. We emphasize that our technique is completely complementary to existing methods for enhanced loading through direct collisional control, and that both methods could be combined to further enhance their performance (see Discussion).

Our protocol is illustrated in Fig. 1(a). In the initialization step, an array of atoms is loaded stochastically and imaged to identify the occupied traps. Atoms are then shelved into a state which is both dark to the loading dynamics and long-lived. Occupied traps are ramped down in intensity through real-time feedback, and the process is restarted by regenerating the MOT or ODT which stochastically reloads the deep, unoccupied traps, while leaving the dark-state atoms in shallow traps unperturbed. The result is a theoretically 25\% higher loading fraction (for an initial loading probability of 50\%); the process can then be repeatedly infinitely for successively higher loading fidelity. We note that iterative loading techniques have been used to demonstrate continuous reloading of dual-atom tweezer arrays, though without increasing the filling fraction of either species~\cite{Singh2022}. The ingredients necessary to implement DSEL are accessible in arrays of alkaline-earth atoms, alkali atoms, or molecules [Fig. 1(b)] (see Discussion).

\textit{Methods.}---The experimental setup we use to implement DSEL with arrays of $^{88}$Sr atoms has been described in detail previously~\cite{Cooper2018,Covey2019A,Madjarov2019}. In brief, atoms are initially captured by a 461 nm (blue) MOT based on the $^1S_0{\leftrightarrow}^1P_1$ transition [Fig. 2(a)]. With a small probability, some atoms can decay from $^1P_1{\rightarrow}^1D_2$~\cite{Cooper2018} and then into either $^3P_2$ or $^3P_1$~\cite{Bauschlicher1985}. Atoms in $^3P_2$ are repumped with 707 nm light through the $^3S_1$ state to $^3P_1$; from there they can decay back to $^1S_0$, closing the MOT cycle. Atoms are typically also repumped out of $^3P_0$ as there is a pathway from $^3P_2{\rightarrow}^3P_0$ during repumping.

Atoms are transferred from the blue MOT to a cold and dense 689 nm (red) MOT based on the narrow $^1S_0{\leftrightarrow} ^3P_1$ transition. From there, an acousto-optic deflector (AOD) generates an arbitrary one-dimensional configuration of optical tweezer traps (813 nm); by tuning the electrical input to the AOD we control the number of tweezers, their spacing, and their relative intensities~\cite{Endres2016}. With a typical depth of 470 $\mu$K, optical tweezers are overlapped with the red MOT, loaded with an ensemble of atoms, and then illuminated with 689 nm light to induce light-assisted collisions~\cite{Zelevinsky2006,Reinaudi2012}. The transitions to the $^1P_1$ and $^3P_1$ states can then be respectively used to image and cool the atoms with high fidelity~\cite{Covey2019A}.

A key feature of the low-lying energy landscape of Sr is the $^3P_0$ state. This state is metastable~\cite{Santra2004}, enabling applications in metrology~\cite{Madjarov2019,Young2020,Bothwell2022,Zheng2022} and it underlies proposals for alkaline-earth based neutral atom quantum computers~\cite{Chen2022,Wu2022}. When in $^3P_0$, atoms are dark to light involved in imaging, cooling, and MOT loading. In Fig. 2(b) we demonstrate this concept directly with three consecutive single-shot fluorescence images. In the first, all atoms are in the absolute ground state, and fluoresce under the 461 nm imaging light; during all images, 689 and 707 nm light is also active for cooling and repumping, respectively. Second, the central atom is locally excited to $^3P_0$, making it dark to the subsequent image. Local addressing is accomplished by lowering the central trap frequency by a factor of two and driving the motional blue sideband with a frequency selective pulse~\cite{Note1}. Finally, all atoms are repumped to the ground state and imaged once more, demonstrating that the $^3P_0$ atoms have survived.

\begin{figure*}[t!]
	\centering
	\includegraphics[width=172mm]{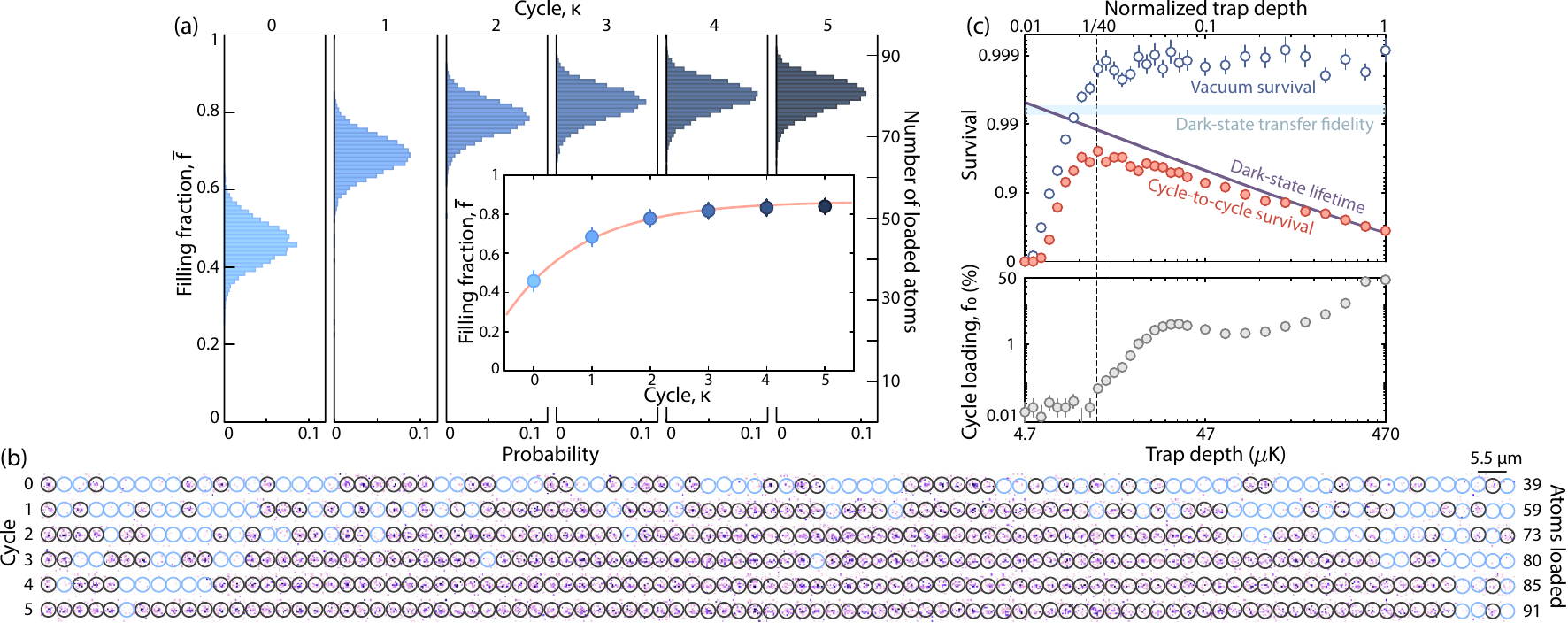}
	\caption{(a) DSEL results in a 95-tweezer array. Shown in the main plot are histograms of the filling fraction of the array after each successive repeated loading cycle. The zeroth cycle is the standard, non-enhanced loading with a loading fraction of 45.88(5)\%, while after five DSEL cycles a maximum of 84.02(4)\% is achieved. Inset: Average and standard deviation of the filling fraction as a function of loading cycle (markers), compared against the prediction from Eq. (1) based on a per-cycle loading fidelity of 0.46, and a cycle-to-cycle survival probability of 0.93 (pink line, see text). (b) Exemplary single-shot fluorescence images from a single set of DSEL cycles, showing the largest-achieved 91 atom array. Sites above the atom-detection threshold are circled in black, while those below the threshold are circled in blue. Note that atoms are additionally rearranged to the middle of the array between each cycle. (c) Cycle-to-cycle survival (red markers) is limited by imperfect dark-state transfer fidelity (blue fill), dark-state lifetime (purple line), and vacuum-limited survival (open markers) during the reloading cycle, here 280 ms. An optimum in survival emerges at a ramp-down factor of 1/40th (a trap depth of 12 $\mu$K). At this trap depth, the probability of reloading an already occupied tweezer is ${\lesssim}0.1$\% (bottom). Data in (c) are taken in a 39-tweezer array with higher fidelity survival and dark-state transfer compared to the 95-tweezer array in (a) and (b).
	} 
	\vspace{-0.5cm}
	\label{fig2}
\end{figure*}

The lifetime of atoms in the $^3P_0$ state is limited by Raman scattering from the high-intensity trapping light~\cite{Dorscher2018}. We find a lifetime of 23(4) s for 12 $\mu$K deep traps while light used for generating the red and blue MOTs is active [Fig. 2(c)] and 0.58(4) s at our typical trap depth of 470 $\mu$K (not shown). At both trap depths the lifetime is unaffected by disabling the MOT light (not shown). In contrast, scattering from the MOT light quickly heats and ejects atoms in the $^1S_0$ state, limiting their trapped lifetime in 12 $\mu$K deep tweezers to less than 2 $\mu$s [Fig. 2(c)]; this amounts to an effective increase in lifetime by a factor of ${\sim}10^7$ when shelving into $^3P_0$.

Isolation from other transitions, in combination with a long lifetime, makes $^3P_0$ an ideal candidate for the dark-state required by DSEL. However, during the standard MOT loading cycle, $^3P_0$ is explicitly repumped. In practice, we find that under our typical conditions, loading without the $^3P_0$ repumper incurs only a 0.7(4)\% drop in loading probability; through optimization of the blue MOT density it is likely that this minor penalty could be further reduced.

\textit{Results.}---We show the results of DSEL for a 95-tweezer array in Fig. 3(a). With an initial loading probability of 45.88(5)\%, after a single cycle of DSEL we achieve 68.43(5)\%, which then saturates to 84.02(4)\% after five cycles of DSEL (the zeroth cycle is the initial standard loading). We show exemplary fluorescence images from a single set of DSEL cycles in Fig. 3(b), demonstrating the cycle-to-cycle growth of array filling, resulting in a maximum of 91 trapped atoms. Note that in Fig. 3(b), atoms are additionally rearranged to the middle of the array between each cycle.

We see the effect of DSEL starts stagnating after around 2 cycles, limited by the probability that a shelved atom survives and still is present after a cycle of DSEL, which we designate as the cycle-to-cycle survival, $p_0$. Assuming the loading process is only controlled by this cycle-to-cycle survival and by the per-cycle loading probability, $f_0$, we predict the total loading probability, $\bar{f}$, after $\kappa$ DSEL cycles is given by \begin{align}
\bar{f}(\kappa)=f_0 \sum\limits_{j=0}^\kappa (p_0-f_0)^j=f_0\frac{1-(p_0-f_0)^{\kappa+1}}{1-(p_0-f_0)\hfill},
\end{align}
where $j$ is a cycle index. For the 95-tweezer array, we find our results are well described by Eq. (1) with $f_0=0.46$ and $p_0=0.93$ [Fig. 3(a), inset].

For a smaller 39-tweezer array with improved atomic survival and dark transfer efficiency, the cycle-to-cycle probability is improved to $p_0\sim0.97$ [Fig. 3(c), upper panel]. We find $p_0$ is primarily controlled by imperfect transfer to $^3P_0$ ~\cite{Note2}, decay of the atoms out of $^3P_0$, and atom survival [Fig. 3(c), upper panel]. This is because if atoms are not transferred to $^3P_0$, or if they decay back to the ground state, then they will be ejected from the traps during the MOT dynamics, leading to loss [Fig. 2(c)]. 

Traps are ramped down immediately following transfer in order to extend the lifetime of atoms in $^3P_0$ and limit this effect. However, traps cannot be ramped down arbitrarily low as the probability for atoms to remain trapped, regardless of electronic state, sharply decreases below a certain trap depth [Fig. 3(c), upper panel] due to their temperature~\cite{Tuchendler2008}. Ultimately we find an optimum of these effects at a trap ramp down factor of 1/40 (a trap depth of 12 $\mu$K); at this trap depth we additionally find the probability of tweezer reloading is ${\lesssim}0.1$\% [Fig. 3(c), lower panel], limiting the probability for new atoms to be loaded into already occupied traps.

\begin{figure}[t!]
	\centering
	\includegraphics[width=86mm]{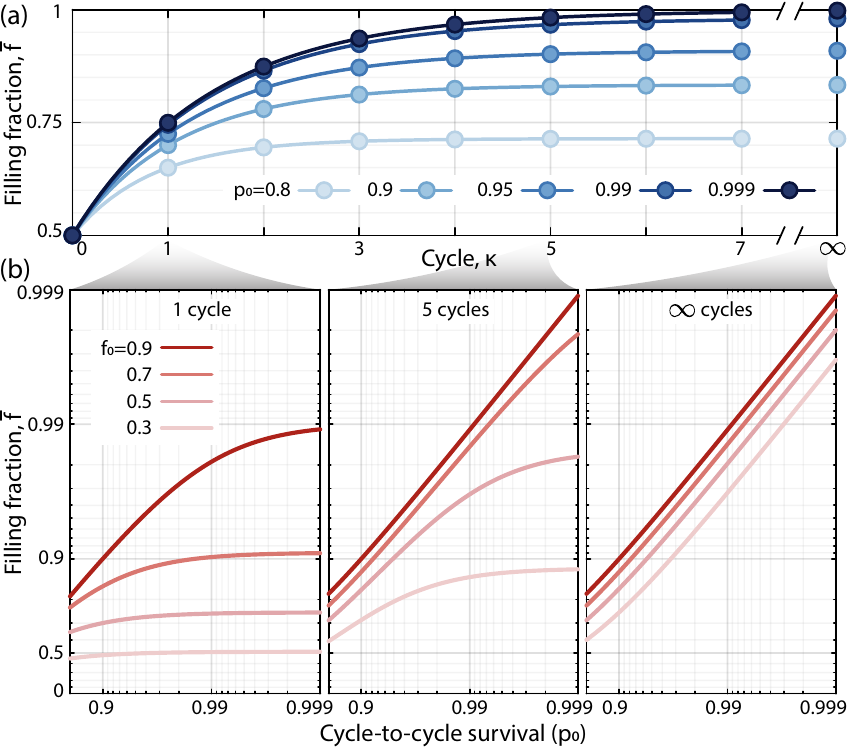}
	\caption{
	(a) Predicted total filling fraction from Eq. (1) with varying number of DSEL cycles, $\kappa$, and a fixed per-cycle filling fraction, $f_0=0.5$. (b) Total filling fraction after DSEL for various cycle numbers, per-cycle filling fractions, $f_0$, and cycle-to-cycle survival probability, $p_0$; $f_0$ values greater than 0.5 are achievable by combining DSEL with existing methods for directly controlling the collisional processes during loading~\cite{Grunzweig2010,Lester2015,Brown2019,Jenkins2022,Aliyu2021,Angonga2022}.
	} 
	\vspace{-0.5cm}
	\label{fig1}
\end{figure}

Having demonstrated DSEL experimentally, in Fig. 4(a) we plot the results of Eq. (1) as a function of total number of cycles for various $p_0$ at a fixed $f_0=0.5$, and include the infinite cycle limit of:
\begin{align}
\bar{f}(\infty)=f_0\frac{1}{1-(p_0-f_0)\hfill}.
\end{align}
We find that the final filling fraction saturates, with the saturation level varying with $p_0$. To understand this behavior, we plot the filling as a function of $p_0$ for 1, 5, and infinite cycles in Fig. 4(b). There we show the results for various $f_0$ ranging from 0.3 to 0.9, given that $f_0$ can be increased with existing methods for enhanced loading based on control over light-assisted collisions~\cite{Grunzweig2010,Lester2015,Brown2019,Jenkins2022,Aliyu2021,Angonga2022}.

In the many cycle limit, we find that DSEL becomes limited primarily by $p_0$; for $p_0$ approaching 1, Eq. (2) reduces at lowest order to $\bar{f}(\infty)\approx1 + (p_0 - 1)/f_0$. With perfect initial loading ($f_0=1$), this further simplifies directly to $\bar{f}(\infty)\approx p_0$. On the other hand, if $p_0=1$ then $\bar{f}(\infty)=1$, no matter the value of $f_0$. Intuitively, this is because the gain from DSEL stems from accurately learning information from atomic imaging to reduce loading entropy. If that information is erroneous, i.e. if $p_0$ is low, then there can be relatively little gain. On the other hand, the gain is maximized when the information gained from imaging the array is most accurate, i.e. when $p_0$ is maximized. With state-of-the-art imaging survival~\cite{Covey2019A}, dark-state transfer~\cite{Schine2022,Madjarov2020}, and optimized MOT loading times, $p_0{>}0.99$ is likely achievable.

\textit{Discussion.}---Though demonstrated here only for the alkaline-earth atom Sr, DSEL is feasible to implement for both alkali and molecular tweezer arrays [Fig. 1(b)]. For the case of laser-coolable molecules, such as CaF, the MOT is typically transferred to an ODT to increase molecule densities~\cite{Anderegg2019}; molecules are then cooled into tweezer potentials via $\Lambda$-cooling on the $|X\ ^2\Sigma^+_{1/2}, N{=}1\rangle{\leftrightarrow}|A\ ^2\Pi_{1/2}, J{=}1/2, +\rangle$ transition. In this case, the dark-state for DSEL could be another long-lived rotational state, such as $N{=}2$, which is far-off resonant from this cooling process.

In the alkali case, such as with Rb, the dark-states would be a long-lived hyperfine manifold such as $F{=}1$; however, because it is difficult to form the initial MOT without repumping from these states, it may be necessary to first transfer from the MOT into an ODT, similar to the molecule case. Atoms can be cooled from the ODT into tweezers using e.g. the $|F=2, mF=2\rangle{\leftrightarrow}|F=3, mF=3\rangle$ transition, to which $F{=}1$ shelved atoms would be dark~\cite{Fuhrmanek2011}. 

Loading directly from an ODT may prove beneficial in general, besides the particular application to the molecular and alkali cases. This is because once the ODT is formed, it can be moved away while the tweezer array is imaged, and then the two can be overlapped once more to perform DSEL without the need to repeat the entire MOT sequence. This could greatly reduce one of the main limitations of DSEL, namely the increased experimental runtime. 

Importantly, DSEL can be implemented with practically no impact on experiment duty cycle by reusing atoms from one iteration of the experiment to the next. Typical experimental sequences end in a stage of readout, which can then be followed by transferring any remaining atoms to the dark-state before preparing the next experimental repetition. This effectively passes atoms from one experimental iteration to the next by implementing a 1-cycle version of DSEL.

If a species has internal structure to its dark-state, coherence could feasibly be maintained within that manifold during DSEL, similar to an approach utilizing dual-species arrays~\cite{Singh2022b}. For instance, the multiple hyperfine sublevels of $^3P_0$ in fermionic alkaline-earth species could be used to shelve atoms while performing mid-circuit replacement of lost atoms~\cite{Cong2022,Wu2022}.

We emphasize that though here we have implemented DSEL in one dimension, the basic principle is applicable to arbitrary-dimensional systems, as long as there is sufficient control over the trap depths of individual traps (or regions of traps) during real-time feedback. This is possible with either crossed AOD, time-multiplexed, or holographic techniques, as are typically used for two-dimensional array generation~\cite{Cooper2018,Nogrette2014,Kim2016,Norcia2018,Saskin2019,Yan2022}.

In summary, we have proposed and demonstrated a protocol for enhancing the filling fraction of an optical tweezer array through dark-state enhanced loading. This protocol achieved a filling fraction of 84.02(4)\% in a 95-tweezer array and yielded an array with 91 atoms; to our knowledge this is the first such enhanced loading demonstrated with a Sr atom array, and the largest one-dimensional array of atoms in optical tweezers to date. Importantly, our protocol is scalable, complementary to existing enhanced loading schemes, and species-agnostic. Ultimately, consistent creation of larger arrays will have benefits for metrology, quantum simulation, and tests of quantum advantage in optical tweezer systems.

\begin{acknowledgements}
\textit{Acknowledgements.}---We acknowledge useful conversations with Joonhee Choi, Xin Xie, Jacob Covey, Hannah Manetsch, and Kon Leung. Further, we thank Arian Jadbabaie, Lo{\"i}c Anderegg, and Yicheng Bao for their insights and suggestions concerning the molecular implementation of our protocol. This material is based upon work supported by the U.S. Department of Energy, Office of Science, National Quantum Information Science Research Centers, Quantum Systems Accelerator. Additional support is acknowledged from the Institute for Quantum Information and Matter, an NSF Physics Frontiers Center (NSF Grant PHY-1733907), the NSF CAREER award (1753386), the AFOSR YIP (FA9550-19-1-0044), the DARPA ONISQ program (W911NF2010021), the Army Research Office MURI program (W911NF2010136), the NSF QLCI program (2016245), and Fred Blum. ALS acknowledges support from the Eddleman Quantum Graduate Fellowship. RF acknowledges support from the Troesh postdoctoral fellowship.
\end{acknowledgements}

% Bibliography
\bibliographystyle{h-physrev}

\end{document}